\documentclass[12pt]{iopart}

\usepackage{iopams}  
\usepackage{graphicx}
\begin{document}

\title[Systematic perturbation approach for a dynamical scaling law 
in a kinetically constrained spin model]
{Systematic perturbation approach for 
a dynamical scaling law in a kinetically constrained spin model}

\author{Hiroki Ohta}

\address{Department of Pure and Applied Sciences,
University of Tokyo, 3-8-1 Komaba Meguro-ku, Tokyo 153-8902, Japan}
\ead{hiroki@jiro.c.u-tokyo.ac.jp}
\begin{abstract}
The dynamical behaviours of a kinetically constrained spin model 
(Fredrickson-Andersen model) on a Bethe lattice are investigated by 
a perturbation analysis that provides exact final states 
above the nonergodic transition point. 
It is observed that 
the time-dependent solutions of the derived dynamical systems 
obtained by the perturbation analysis become 
systematically closer to the results obtained by Monte Carlo simulations 
as the order of a perturbation series is increased.
This systematic perturbation analysis also clarifies the existence of a dynamical scaling law, which provides a implication for a universal relation
between a size scale and a time scale near the nonergodic transition. 
\end{abstract}

\pacs{05.50.+q, 64.70.Q-, 64.70.qj}
\maketitle

\newcommand{\bra}{\left\langle}
\newcommand{\ket}{\right\rangle}
\newcommand{\pder}[2]{\dfrac{\partial #1}{\partial  #2}}
\newcommand{\pdert}[2]{\dfrac{\partial^2 #1}{\partial  #2^2}}
\newcommand{\der}[2]{\dfrac{d #1}{d  #2}}
\newcommand{\bv}[1]{{\boldsymbol #1}}
\newcommand{\p}{\partial_t}
\newcommand{\ovl}[1]{\overline #1}

\newcommand{\hc}{h_{\rm c}}
\newcommand{\rhov}{\rho_{\rm v}}
\newcommand{\nug}{\nu_{\rm g}}
\newcommand{\num}{\nu_{\rm m}}
\newcommand{\Nc}{N_{\rm c}}
\newcommand{\Dt}{\Delta t }
\newcommand{\thetas}{\theta^{\rm s}}
\newcommand{\Domh}{{\rm Dom}_h}
\newcommand{\rhoq}[1]{\rho_{#1}^{\rm q}}

\newcommand{\du}{d_{\rm u}}
\newcommand{\Rcsp}{R_{\rm c}^{\rm sp}}
\newcommand{\Rceq}{R_{\rm c}^{\rm eq}}
\newcommand{\rhos}{{\boldsymbol  \rho}_{\rm c}}
\newcommand{\ep}{\epsilon}

\section{Introduction}
Recently, soft materials such as colloidal and granular systems have 
attracted considerable interest owing to their rich behaviours. 
For instance, such systems can be in a supercooled state under conditions 
of low temperatures and high densities \cite{Cavagna}. 
Under such conditions, a characteristic time acts
as a function of system parameters such as temperature or density, 
and often obeys a non-Arrhenius law, 
which substantially influences the properties of the materials. 
Understanding the mechanism of such anomalous dynamical behaviours in 
many-body systems
is important in the field of statistical physics.

Kinetically constrained spin model (KCSM) 
is a simple model that follows the non-Arrhenius law \cite{FA,KA,Sollich}. 
Thus far, it has been rigorously proved that the characteristic times
in some kinds of KCSM on finite-dimensional lattices 
show super-Arrhenius type and Vogel-Fulcher type behaviours \cite{Toninelli0,Toninelli1}.   
Furthermore, in the case of a KCSM on a Bethe lattice, another non-Arrhenius type behaviour 
of a characteristic time has been found by Monte Carlo simulations \cite{Toninelli2}. 
From a static aspect, this non-Arrhenius type behaviour is 
due to a nonergodic transition corresponding to $k$-core percolation; 
this transition is not a thermodynamic phase transition.
Concretely, at this percolation point, 
the characteristic time diverges with a power law 
(non-Arrhenius law); this divergence 
is supposedly controlled by a mode-coupling equation \cite{Toninelli2}. 

The above mentioned mode-coupling equations are also believed 
to be related with the anomalous dynamical behaviours of colloidal 
or granular systems. 
A related conjecture is that a finite-dimensional system mimics
a nonergodic transition described by a mode-coupling equation 
in a mean field sense although 
it is not a true nonergodic transition 
but a strong finite-size effect \cite{Cavagna}. 
In the case of KCSM, it has been rigorously proved 
that the nonergodic transition observed specifically
in a KCSM on a Bethe lattice does not occur in the model on 
a finite dimensional lattice although there are 
the strong finite-size effects arising from 
the nonergodic transition on the Bethe lattice \cite{Toninelli3,Toninelli4}.

This leads us to consider 
whether the mechanisms of the appearance of such strong finite-size effects 
for different finite-dimensional systems have common features.
In order to resolve this problem, 
it is necessary to find a relationship 
between KCSM and mode-coupling equations. 
However, the mode-coupling equation describing 
the nonergodic transition has not been derived yet for KCSM on Bethe lattices: 
there have been some related studies on the derivations of 
mode-coupling equations for KCSM 
\cite{Eisinger,Kawasaki,Andersen1,Andersen2,Szamel}.

In this paper, as a preliminary step to understand such a relationship,
we attempt to clarify the dynamical aspect 
of the universality class of the nonergodic transition observed 
in Fredrickson-Andersen model (a KCSM) on a Bethe lattice. 
Concretely, we derive approximately dynamical systems from 
this model using a perturbation analysis which provides exact final states 
as stationary solutions above the nonergodic transition.
We find that the universal class of the nonergodic 
transition cannot be captured by each dynamical systems even at any order by itself.
Nevertheless, we find that the differences between the time-dependent 
solutions of the derived dynamical systems and 
the results obtained by Monte Carlo (MC) simulations are 
systematically reduced on a perturbation series.
Furthermore, we find that this systematic perturbation analysis 
clarifies the existence of a dynamical scaling law, 
which provides a implication for 
a universal relation between a size scale and 
a time scale near the nonergodic transition.

\section{Model}
Let us consider a regular random graph consisting of 
$N\in \mathbb{N}$ sites, each of which connects to $c\in \mathbb{N}$ 
sites chosen randomly, where $\mathbb{N}$ is the set of natural numbers.
Then ${\rm G}(c,N)$ is defined as a set of such regular random graphs.
For the spin variable $\sigma_i \in \{-1,1\}$ 
defined on each site $i\in \{1,\cdots,N\}$ 
in a graph $\mathcal{G}\in{\rm G}(c,N)$, 
the Hamiltonian we consider is
\begin{eqnarray}
H(\bv{\sigma})= \frac{1}{2}\sum_{i=1}^N \sigma_i,
\end{eqnarray} where we express $\bv{\sigma}\equiv(\sigma_i)_{i=1}^N$ collectively.
Here, as a preliminary step to define the dynamics of the system, 
let us consider a transition rate 
$r(\bv{\sigma},F_i\bv{\sigma})$ from $\bv{\sigma}$ to $F_i\bv{\sigma}$, 
which satisfies the detailed balance condition. 
Here, $F_i$ is the spin flip operator such that $F_i\bv{\sigma}=(\sigma_1,\cdots,-\sigma_i,\cdots,\sigma_N)$. Let $B_i$ be a set of sites connected to site $i$.
Next, we consider the following dynamical rule. 
If the number of upward spins on the sites in set $B_i$ are 
more than or equal to $k\in \mathbb{N}$, 
the spin on site $i$ does not flip absolutely;
otherwise, the spin on site $i$ flips 
at a transition rate $r(\bv{\sigma},F_i\bv{\sigma})$.
In other words, under this rule, 
the transition rate $T(\bv{\sigma}\to F_i\bv{\sigma})$ from 
$\bv{\sigma}$ to $F_i\bv{\sigma}$ is expressed by 
$r(\bv{\sigma},F_i\bv{\sigma})\Theta( 2k - c - \sum_{j\in B_i}\sigma_j)$, 
where $\Theta(x)=1$ for $x > 0$, otherwise $0$. 
We define the situation of a spin $\sigma_i$ 
with $\Theta(2k-c -\sum_{j\in B_i}\sigma_j)=0$ 
as `kinetically constrained' or simply `constrained'.
The master equation for the probability $P(\bv{\sigma},t)$ that 
spin configuration at time $t$ is $\bv{\sigma}$ is 
\begin{eqnarray}
\partial_t P(\bv{\sigma},t)=\sum_{i=1}^N[T(F_i\bv{\sigma}\to\bv{\sigma})
P(F_i\bv{\sigma},t)-T(\bv{\sigma}\to F_i\bv{\sigma})P(\bv{\sigma},t)].
\label{mas}
\end{eqnarray} 
In this paper, we consider the case
\begin{eqnarray}
r(\bv{\sigma},F_i\bv{\sigma})=\min(1,\exp(\frac{H(\bv{\sigma})
-H(F_i\bv{\sigma}))}{T}).
\end{eqnarray} 
Under this constrained dynamical rule, it may be confirmed that the canonical distribution 
is a stationary distribution. 
Further, in equilibrium, 
the magnetization per site is $m_{\rm eq}(T)=\tanh(1/2T)$, 
and the energy density is $-\tanh(1/2T)/2$.
Therefore, there are no thermodynamic phase transitions in the system.
In this paper, MC simulations are performed by the following rule.
First, a site $i$ is randomly chosen. Next, the spin on site $i$ flips with 
the probability $T(\bv{\sigma}\to F_i\bv{\sigma})$. This step is 
repeated and time $t=1$ is defined by $N$ repeated steps.
It is plausible that in the thermodynamic limit, this MC simulation 
is the same as the dynamics of the system described by equation (\ref{mas}).

\begin{figure}
\includegraphics[width=8cm,clip]{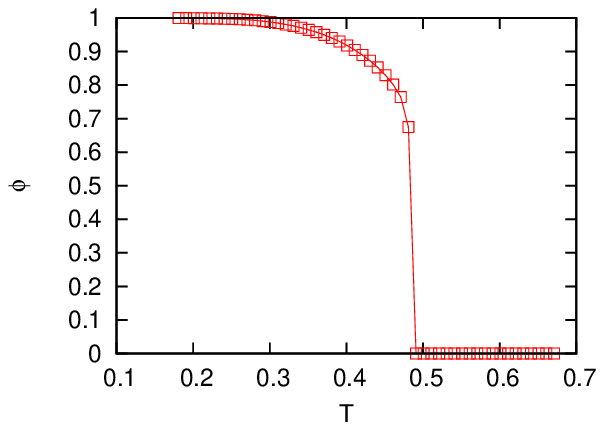}\includegraphics[width=6cm,clip]{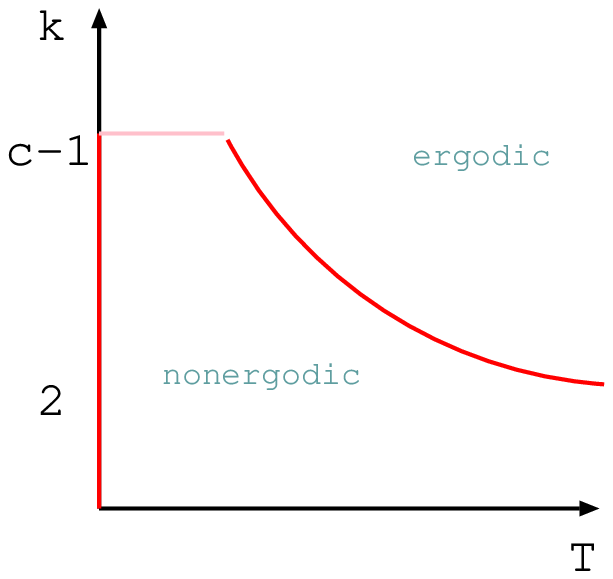}
\caption{(left) The fraction $\phi$ of permanently constrained spins for 
$c=4,k=3$.
(right) Schematic phase diagram.}
\label{static}
\end{figure}

Here, we briefly review the static aspect of a nonergodic transition 
in the system caused by the constrained dynamics, 
which is discussed in the previous study \cite{Toninelli2}.
Suppose that a spin $\sigma_i$ is constrained.
If this constraint is permanent, 
we define the situation of a spin $\sigma_i$ as `frozen'.
Here, let us consider a Cayley tree, which has the same local structures as 
those of the random graph, ignoring the effects of the loop length $O(\log N)$. 
Let $g\in\{1,2,\cdots,g_{\rm max}\}$ 
be a generation of a Cayley tree where $g=1$ is assigned to the root.
Let us consider the probability $Q_g$ that a spin at 
the $g$-th generation obtained under equilibrium conditions dependent on $T$ 
is frozen and upward without considering the state of spin at 
the $(g-1)$-th generation. 
From the tree structure of the graph, we can obtain the relation 
\begin{eqnarray}
Q_{g-1}=F(Q_g),\\
F(Q_g)=p(T)\sum_{n=k}^{c-1}
\left(
\begin{array}{c}
c-1 \\ n
\end{array}
\right)
Q_g^{n}(1-Q_g)^{c-1-n}, \label{rec}
\end{eqnarray} where $p(T)\equiv 1/(1+\exp(-1/T))$.
It should be noted that by solving recursion equation (\ref{rec}) 
for given values of $Q_{g_{\rm max}}$,
$Q_g$ for $g\ll g_{\rm max}$ becomes 
a solution $Q(h)$ satisfying $Q(T)=F(Q(T))$.
When $c=4$ and $k=3$, $Q(T)$ 
is zero for sufficiently high temperatures.  
However, when the temperature is decreased,
 $Q(T)$ suddenly can take a finite value at finite temperature 
$T_{\rm c}=0.480898$, 
as shown in the left-hand side of figure \ref{static}. 
This singular point is $k$-core percolation point, 
below which the system is nonergodic.
Using the quantity $Q(T)$, 
the fraction $\phi$ of frozen spins is described as 
\begin{eqnarray}
\phi(T)=p(T)\sum_{n=k}^c
\left(
\begin{array}{c}
c \\ n
\end{array}
\right)
Q(T)^n(1-Q(T))^{c-n}\nonumber\\+ (1-p(T))\sum_{n=k}^c
\left(
\begin{array}{c}
c \\ n
\end{array}
\right)
(Q'(T))^n(1-Q'(T))^{c-n}, 
\end{eqnarray} where 
$Q'(T)=p(T)\sum_{n=k}^{c-1}\left(
\begin{array}{c}
c-1 \\ n
\end{array}
\right)Q(T)^n(1-Q(T))^{c-1-n}$.
In this model, it has been known that for $2<k<c$, 
this type of nonergodic transition occurs at $T=T_c$ with $0<T_c<\infty$. 
For $k=c$, $T_{\rm c}$ is zero, and for $k\le 2$, $T_{\rm c}$ is $\infty$.  
The schematic phase diagram is shown in the right-hand side of figure \ref{static}.

\section{Simple analysis of the dynamics} 
Although master equation (\ref{mas}) provides the complete information 
about the system, 
it is very difficult to extract useful information of the system 
from (\ref{mas}) because the system has 
$2^N$ states, which is quite a large number when $N$ is large. 
To avoid this difficulty, 
we consider describing the system by $N_{\rm eff}\in \mathbb{N}$ 
number of variables
and derive approximately a dynamical system closed by the set of the variables, 
where $N_{\rm eff}$ remains finite when $N\to\infty$. 
For simplicity, we consider the relaxation behaviours of the system 
for the initial condition $\hat{m}(0)=-1$ where $\hat{m}(t)\equiv\sum_{i=1}^N\sigma_i/N$. 
It should be noted that the spin configuration for $\hat{m}(0)=-1$ 
has no constrained spins.
In the following analysis, we fix a graph $\mathcal{G}\in{\rm G}(c,N)$ 
for sufficiently large $N$ without considering ensembles for ${\rm G}(c,N)$.
In other words, the following analysis can be applicable to almost all graphs
$\mathcal{G}\in {\rm G}(c,N)$ in the thermodynamic limit.

As a first step to derive an effective dynamical system, 
let $P_i(\sigma;t)$ be the probability that $\sigma_i$ takes $\sigma$ at time $t$
and $P_i((\sigma_j')|\sigma';t)$ 
be the probability that the spin configuration on the sites in set $B_i$ is 
$(\sigma_j')_{j\in B_i}$ provided that $\sigma_i$ takes $\sigma'$ at time $t$.
Then, we have the following exact evolution equation.
\begin{eqnarray}
&\partial_t P_i(\sigma;t) = \sum_{(\sigma_{j}')}
\Theta(2k-c-\sum_{j'\in B_i}\sigma_{j'}')\nonumber\\
&[- r_{\sigma}P_i(\sigma;t)P_i((\sigma_{j}')|\sigma;t)
+ r_{-\sigma}P_i(-\sigma;t)P_i((\sigma_{j}')|-\sigma;t)], \label{eq0}
\end{eqnarray} where we define $r_{\sigma}\equiv\min(1,\exp(-\sigma/T))$.
Here, we assume that the value of $P_{i}((\sigma_j')|\sigma)$ 
does not depend on the chosen site $i$. 
This assumption may be plausible because at least, in this case, 
inhomogeneous properties of the system arising from 
the effects of loops of the random graph $\mathcal{G}$ may be 
negligible in the thermodynamic limit.
This assumption corresponds to the assumption that 
$P_{i}((\sigma_j')|\sigma)$ is the same as the conditional probability 
$P((\sigma_j')|\sigma;t)$ 
that if a site with spin variable $\sigma$ is randomly chosen, 
the spin configuration of its nearest neighbor sites is $(\sigma_j')$. 
With this assumption, equation (\ref{eq0}) is rewritten as
\begin{eqnarray}
&\partial_t \rho_\sigma(t) = \sum_{(\sigma_{j}')}
\Theta(2k-c-\sum_{j'\in B_\sigma}\sigma_{j'}')\nonumber\\
&[- r_{\sigma}\rho_{\sigma}(t)P((\sigma_{j}')|\sigma;t)
+ r_{-\sigma}\rho_{-\sigma}(t)P((\sigma_{j}')|-\sigma;t)], \label{eq01}
\end{eqnarray} where $B_\sigma$ is a set of sites connected to a 
site with spin variable $\sigma$ and $\rho_\sigma\equiv\sum_{i=1}^NP_i(\sigma)/N$. 
It is plausible that $\rho_\sigma$ is identical to $\sum_i \delta(\sigma-\sigma_i)/N$ 
in the thermodynamic limit $N\to\infty$. 
That is, the magnetization is expressed as 
$m(t)=\sum_{\sigma}\sigma \rho_{\sigma}(t)$.

Next, we use the approximation that 
the spin variables on individual sites are independent of each other, 
which is exact for equilibrium spin configurations.
That is, we rewrite $P((\sigma_{j}')|\sigma;t)$ as
\begin{eqnarray}
P((\sigma_{j}')|\sigma;t)= \prod_{j\in B_{\sigma}}\rho_{\sigma_{j}'}(t).\label{rep0}
\end{eqnarray} 
Further, we can obtain a simpler expression as follows.
\begin{eqnarray}
\sum_{(\sigma_{j}')}\Theta(c-2k-\sum_{j'\in B_\sigma}\sigma_{j'}')
\prod_{j\in B_\sigma}\rho_{\sigma_{j}'}(t)= \sum_{l=f}^c\left(
\begin{array}{c}
c \\ l
\end{array}
\right)\rho_{-1}(t)^l\rho_{+1}(t)^{c-l},\label{cal0}
\end{eqnarray} where we define $f\equiv c-k+1$.
Equations (\ref{eq0}), (\ref{rep0}), and (\ref{cal0}) lead to 
a closed dynamical system in terms of $\bv{\rho}^{(0)}\equiv(\rho_\sigma)_\sigma$, 
\begin{eqnarray}
\partial_t \bv{\rho}^{(0)} = \bv{G}^{(0)}(\bv{\rho}^{(0)}).\label{dyn0}
\end{eqnarray} 

\begin{figure}
\centering
\includegraphics[width=8cm,clip]{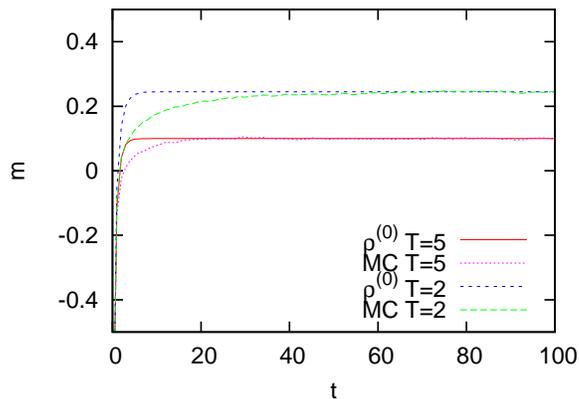}
\caption{Time-dependent magnetization $\sum_{\sigma}\sigma \rho_\sigma$ 
described by (\ref{dyn0}) and $\hat{m}$ 
by the MC simulations with $N=10^6$ for $c=4,k=3$.}
\label{met0}
\end{figure}
It should be noted that stationary solutions of dynamical system (\ref{dyn0}) 
provide the exact final states above the nonergodic transition point. 
However, behaviours of the system 
at intermediate time scales are very different 
from the MC simulations, as seen in figure \ref{met0}.
This means that approximation (\ref{rep0}) 
fails to capture heterogeneous spin configurations responsible for 
the dynamics in intermediate time scales. 
In this paper, we use the fourth-order Runge-Kutta method for 
obtaining solutions of dynamical systems with the time discretization $dt=10^{-2}$. 

Here, let us consider a persistent time $\tau_p(i)$ of a site $i$, 
which is the time span for a spin to flip at site $i$.
Next, let us focus on the target site $i$ 
in the set ${\rm C}_i$ of sites on which constrained spins are connected to each other.
Clearly, $\tau_p(i)$ strongly depends on the entire spin configuration 
of sites in set ${\rm C}_i$ because the spins from a edge site in set ${\rm C}_i$ 
must be flipped in order to flip the target spin $\sigma_i$.
Furthermore, the spin configuration on the sites in set ${\rm C}_i$ 
is heterogeneous in terms of the spin direction 
if it is prepared from the equilibrium spin configurations.
This suggests that in order to detect growing relaxation times of the system 
with respect to the persistent time, 
it is necessary to obtain information 
of the heterogeneous configurations of the connected constrained spins.

\section{Perturbation analysis of the dynamics}
From the results in the previous section, 
heterogeneous spin configurations seem to play 
important roles in the growing relaxation times of the system. 
Therefore, in order to continue our analysis,
we attempt to apply the information of the surrounding 
spin configurations of a target site perturbatively to effective variables 
by increasing the value of $N_{\rm eff}$.
Such a method has been previously applied to some systems
and was successful in determining some nontrivial dynamical properties 
\cite{Semerjian,Ohta2}. 

\subsection{First layer}
We attempt to apply the information of the first `layer' 
of surrounding spin configurations 
of a target site to effective variables. 
To this end, we suppose $w_i\in\{1,\cdots,c\}$ as the number of downward spins 
on sites in set $B_i$.
With this representation, site $i$ is characterized by $(\sigma_i,w_i)$. 
Here, let $P_i(\sigma,w;t)$ be the probability that 
$(\sigma_i,w_i)$ takes $(\sigma,w)$
and $P_{ij}(\sigma,w|\sigma',w';t)$ be the conditional probability
that $(\sigma_i,w_i)$ takes $(\sigma,w)$ 
provided that $(\sigma_j,w_j)$ takes $(\sigma',w')$.
Of course, a trivial relation $\sum_{\sigma}\sum_{w=0}^cP_i(\sigma,w;t)=1$ holds.
Using these expressions, we have the following exact evolution equation.
\begin{eqnarray}
\partial_t P_i(\sigma,w;t) = (-r_{\sigma}P_i(\sigma,w;t)
+r_{\sigma}P_i(\sigma,w;t))\ovl{\Theta}(w-f) \nonumber\\
+ \sum_{j \in B_i}\sum_{\sigma'}\sum_{w'= f}^cr_{\sigma'}P_j(\sigma',w';t)\nonumber\\
(P_{ij}(\sigma,w+\sigma'|\sigma',w';t)-P_{ij}(\sigma,w|\sigma',w';t)),
\end{eqnarray} where $f$ is defined as $c-k+1$ and $\ovl{\Theta}(x)=1$ for $x\ge 0$, otherwise 0.

Here, we assume that the value of $P_{ij}(\sigma,w|\sigma',w')$ does not depend 
on the chosen sites $i$ and $j$ 
if two sites $i$ and $j$ are chosen among the pairs of sites which have
the same distance. 
This assumption may be plausible because at least, in this case, 
inhomogeneous properties of the system arising from 
the effects of loops of the random graph $\mathcal{G}$ may be 
negligible in the thermodynamic limit.
This assumption corresponds to the assumption that $P_{ij}(\sigma,w|\sigma',w';t)$ 
with $j\in B_i$ is the same as the conditional probability 
$P_{{\rm c}1}(\sigma,w|\sigma',w';t)$ 
that after a site characterized by $(\sigma',w')$ is randomly chosen 
, then one of its nearest neighbor sites, when randomly chosen, 
is characterized by $(\sigma,w)$. 

Here, let us define 
\begin{eqnarray}
P_{{\rm c}1}^{\sigma}(w|\sigma',w';t)\equiv 
\frac{P_{{\rm c}1}(\sigma,w|\sigma',w';t)}{P_{{\rm c}1}(\sigma|\sigma',w';t)},
\end{eqnarray} where $P_{{\rm c}1}(\sigma|\sigma',w';t)$ is defined 
in a similar way as $P_{{\rm c}1}(\sigma,w|\sigma',w';t)$.
In fact, we can obtain
\begin{eqnarray}
P_{{\rm c}1}(\sigma|\sigma',w';t) = \frac{1}{c}C_\sigma(w'),\\
C_\sigma(w')=\left\{
\begin{array}{ll}
c-w' & (\sigma=1)\\
w' & (\sigma=-1).
\end{array}\right.
\end{eqnarray}
Therefore, 
\begin{eqnarray}
\partial_t \rho_{\sigma,u}(t) = (-\sigma r_{\sigma}\rho_{\sigma,w}(t)
+\sigma r_{\sigma}\rho_{\sigma,w}(t))\ovl{\Theta}(u-f) \nonumber\\
+ \sum_{\sigma'}\sum_{w'=f}^cr_{\sigma'}\rho_{\sigma',w'}(t)C_\sigma(w')
(P_{{\rm c}1}^{\sigma}(w+\sigma'|\sigma',w';t)
-P_{{\rm c}1}^{\sigma}(w|\sigma',w';t)),\label{eq1}
\end{eqnarray} where $\rho_{\sigma,w}=\sum_{i=1}^NP_i(\sigma,w)/N$. 
It is plausible that $\rho_{\sigma,w}$ corresponds to 
$\sum_{i=1}^N\delta(\sigma-\sigma_i)\delta(w-w_i)/N$ in the thermodynamic limit $N\to\infty$. 
That is, the magnetization is expressed as 
$m(t)=\sum_{\sigma}\sum_{w=0}^c\sigma \rho_{\sigma,w}(t)$.
Here, in order to obtain a closed description 
in terms of $\bv{\rho}^{(1)}\equiv(\rho_{\sigma,w})_{\sigma,w}$, 
we use the following approximation.
\begin{eqnarray}
P_{{\rm c}1}^{\sigma}(w|\sigma',w';t)=P_{{\rm c}1}^{\sigma}(w|\sigma';t) \label{app1},
\end{eqnarray} which is exact for equilibrium states.
Hence, we can obtain
\begin{eqnarray}
P_{{\rm c}1}^{\sigma}(w|-1;t)&=\frac{w\rho_{\sigma,w}(t)}{\sum_{w=0}^cw\rho_{\sigma,w}(t)},\label{cal11}\\
P_{{\rm c}1}^{\sigma}(w|+1;t)&=\frac{(c-w)\rho_{\sigma,w}(t)}{\sum_{w=0}^c(c-w)\rho_{\sigma,w}(t)}.\label{cal12}
\end{eqnarray}
Equations (\ref{eq1}), (\ref{app1}), (\ref{cal11}) and (\ref{cal12}) lead to 
a closed dynamical system in terms of 
$\bv{\rho}^{(1)}$ as follows.
\begin{eqnarray}
\partial_t \bv{\rho}^{(1)} = \bv{G}^{(1)}(\bv{\rho}^{(1)}). \label{dyn1}
\end{eqnarray}

\begin{figure}
\centering
\includegraphics[width=8cm,clip]{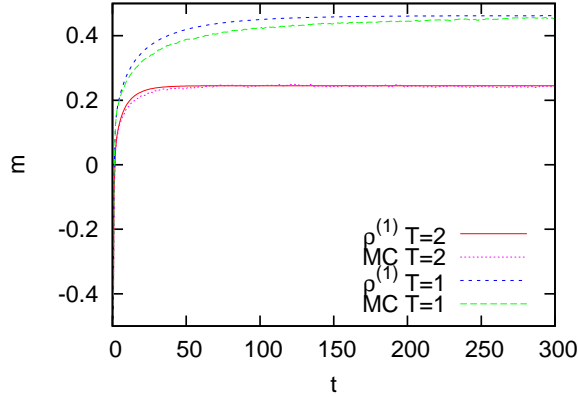}
\caption{Time-dependent magnetization $\sum_{\sigma}\sum_{w=0}^c\sigma \rho_{\sigma,w}$ 
described by (\ref{dyn1}) and $\hat{m}$ by
the MC simulations with $N=10^6$ for $c=4,k=3$.}
\label{met1}
\end{figure}
The stationary solutions of dynamical system (\ref{dyn1}) 
also provide the exact final states above the nonergodic transition point.
The dynamical behaviours of dynamical system (\ref{dyn1}) 
are closer to the MC simulations than those of dynamical system (\ref{dyn0}). 
However, as seen in figure \ref{met1}, the behaviours of dynamical system (\ref{dyn1})
are also gradually deviated by the MC simulations 
if the temperature approaches the transition point $T_{\rm c}$.  
This deviation indicates that 
approximation (\ref{app1}) does not determine 
the behaviours at low temperatures. 

\subsection{Second layer}
We attempt to apply the information 
of the second `layer' of the surrounding spin configurations 
of a target site to effective variables.
Here, for including such effects, 
we consider the following characterization of a site.
First, let $l_i\in\{1,\cdots,c\}$ be the number of downward spins on sites in set $B_i$. 
Second, let $u_i\in \{1,\cdots,c\}$ be 
the number of upward-constrained spins 
and $v_i\in \{1,\cdots,c\}$ be 
the number of downward constrained spins on the sites in set $B_i$. 
Thus, a site $i$ is characterized by $(\sigma_i,l_i,u_i,v_i)$.
With these expressions, let $P_i(\sigma,l,u,v;t)$ be the probability 
that $(\sigma_i,l_i,u_i,v_i)$ takes $(\sigma,l,u,v)$. 
In addition, let $P_{ij}(\sigma,l,u,v|\sigma',l',u',v';t)$ 
be the conditional probability that 
$(\sigma_i,l_i,u_i,v_i)$ takes $(\sigma,l,u,v)$
provided that $(\sigma_j,l_j,u_j,v_j)$ takes $(\sigma',l',u',v')$, 
and let $P_{i_1i_2i_3}((\sigma_1,l_1,u_1,v_1),(\sigma_2,l_2,u_2,v_2)|\sigma_3,l_3,u_3,v_3;t)$ 
be the conditional probability that 
$(\sigma_{i_1},l_{i_1},u_{i_1},v_{i_1})$ and $(\sigma_{i_2},l_{i_2},u_{i_2},v_{i_2})$ 
take $(\sigma_1,l_1,u_1,v_1)$ and $(\sigma_2,l_2,u_2,v_2)$, respectively,
provided that $(\sigma_{i_3},l_{i_3},u_{i_3},v_{i_3})$ takes $(\sigma_3,l_3,u_3,v_3)$.
Of course, a trivial relation
$\sum_{\sigma}\sum_{l=0}^c\sum_{u=0}^{c-l}\sum_{v=0}^{l}P_i(\sigma,l,u,v;t)=1$ 
holds .

As in the previous sections, 
we assume that the value of $P_{ij}(\sigma,l,u,v|\sigma',l',u',v')$ 
does not depend on the chosen sites $i,j$ 
if two sites $i,j$ are chosen among the pairs of sites which have the same distance. 
In addition, we assume that the value of 
$P_{i_1i_2i_3}((\sigma_1,l_1,u_1,v_1),(\sigma_2,l_2,u_2,v_2)|\sigma_3,l_3,u_3,v_3)$ 
does not depend on the chosen sites $i_1,i_2,i_3$ 
if three sites  $i_1$, $i_2$ and $i_3$ are chosen among the sets of three sites 
which have the same relationship for their distances in the order. 
These assumptions are
plausible because at least, in this case, 
inhomogeneous properties of the system arising from the
effects of loops of the random graph $\mathcal{G}$ 
may be negligible in the thermodynamic limit.
This assumption corresponds to the assumption that 
$P_{ij}(\sigma,l,u,v|\sigma',l',u',v')$, where $j\in B_{i}$, 
is the same as $P_{{\rm c}1}(\sigma,l,u,v|\sigma',l',u',v')$
and to the assumption that 
$P_{i_1i_2i_3}((\sigma_1,l_1,u_1,v_1),(\sigma_2,l_2,u_2,v_2)|\sigma_3,l_3,u_3,v_3)$, 
where $i_2\in B_{i_1}, i_3\in B_{i_2}(i_3\neq i_1)$, 
is be the conditional probability 
$P_{{\rm c}2}((\sigma_1,l_1,u_1,v_1),(\sigma_2,l_2,u_2,v_2)|\sigma_3,l_3,u_3,v_3)$ 
that a site characterized by $(\sigma_3,l_3,u_3,v_3)$ is randomly chosen first,
after which one of two connected sites, 
which is connected to the first chosen site, 
is randomly chosen, then among the two connected sites, 
the far site from the first chosen site 
is characterized by $(\sigma_1,l_1,u_1,v_1)$, 
and the near site is characterized by $(\sigma_2,l_2,u_2,v_2)$.

Further, we define 
\begin{eqnarray}
P_{{\rm c}1}^{l,\sigma}(u,v|\sigma',l',u',v')\equiv 
\frac{P_{{\rm c}1}(\sigma,l,u,v|\sigma',l',u',v')}
{P_{{\rm c}1}(\sigma,l|\sigma',l',u',v')}.
\end{eqnarray}
Using this representation, we also define
\begin{eqnarray}
P_{{\rm c}2}^{l_1,\sigma_1}(l_1,u_1,v_1|(\sigma_2,l_2,u_2,v_2),(\sigma_3,l_3,u_3,v_3))\equiv\nonumber\\
\frac{P_{{\rm c}2}((\sigma_1,l_1,u_1,v_1),(\sigma_2,l_2,u_2,v_2)|\sigma_3,l_3,u_3,v_3)}
{P_{{\rm c}2}(\sigma_1,l_1|(\sigma_2,l_2,u_2,v_2),(\sigma_3,l_3,u_3,v_3))}\nonumber\\
P_{{\rm c}1}^{l_2,\sigma_2}(u_2,v_2|\sigma_3,l_3,u_3,v_3)
P_{{\rm c}1}(\sigma_2,l_2|\sigma_3,l_3,u_3,v_3).
\end{eqnarray}
In fact, we can obtain
\begin{eqnarray}
P_{{\rm c}1}(\sigma,l|\sigma',l',u',v')=
\frac{1}{c}C_{\sigma,l}(l',u',v'),\nonumber\\
C_{\sigma,l}(l',u',v')=\left\{
\begin{array}{ll}
u' & (\sigma=1, l < f)\\
v' & (\sigma=-1, l < f)\\
l'-v' & (\sigma=-1, l \ge f)\\
c-l'-u' & (\sigma=1, l \ge f),
\end{array}\right.\\
P_{{\rm c}2}(\sigma_1,l_1|(\sigma_2,l_2,u_2,v_2),(\sigma_3,l_3,u_3,v_3))
=\frac{1}{c-1}C_{\sigma_1,l_1}(l_2,u_2,v_2,\sigma_3,l_3),\nonumber\\
C_{\sigma_1,l_1}(l_2,u_2,v_2,\sigma_3,l_3)=\nonumber\\
\left\{
\begin{array}{ll}
u_2 - \delta(\sigma_3-\sigma_1)\Theta(f-l_3) & (\sigma_1=1, l_1 < f)\\
v_2 - \delta(\sigma_3-\sigma_1)\Theta(f-l_3) & (\sigma_1=-1, l_1 < f)\\
l_2-v_2 - \delta(\sigma_3-\sigma_1)\ovl{\Theta}(l_3-f) & (\sigma_1=-1, l_1 \ge f)\\
c-l_2-u_2 - \delta(\sigma_3-\sigma_1)\ovl{\Theta}(l_3-f) & (\sigma_1=1, l_1 \ge f).
\end{array}\right.
\end{eqnarray}
Using these expressions, we can write the evolution equation as follows.
\begin{eqnarray}
&\partial_t \rho_{\sigma,l,u,v}= (-r_{\sigma}\rho_{\sigma,l,u,v}+r_{-\sigma}\rho_{-\sigma,l,u,v})
\ovl{\Theta}(l-f) \nonumber\\
&+ \sum_{\sigma'}\sum_{l'=f}^c\sum_{u'=0}^{c-l'}\sum_{v'=0}^{l'}r_{\sigma'}\rho_{\sigma',l',u',v'}
C_{\sigma,l}(l',u',v')\nonumber\\
&[P_{{\rm c}1}^{l,l+\sigma'}(u,v|\sigma',l',u',v')-P_{{\rm c}1}^{l,\sigma}(l,u,v|\sigma',l',u',v')]\nonumber\\
&+ \sum_{l''=f}^c\sum_{u''=0}^{c-l''}\sum_{v''=0}^{l''}r_{+1}\rho_{+1,l'',u'',v''}
C_{\sigma',l'}(l'',u'',v'')\nonumber\\
&\sum_{\sigma'}\sum_{u'=0}^{c-f+1}\sum_{v'=0}^{f-1}P_{{\rm c}1}^{f-1,\sigma'}(u',v'|+1,l'',u'',v'')
C_{\sigma,l}(l',u',v',+1,l'')\nonumber\\
&[P_{{\rm c}2}^{l,\sigma}(u+\delta(\sigma'+1),v+\delta(\sigma'-1)|(\sigma',f-1,u',v'),(+1,l'',u'',v''))\nonumber\\
&-P_{{\rm c}2}^{l,\sigma}(u,v|(\sigma',f-1,u',v'),(\sigma'',l'',u'',v''))]\nonumber\\
&+\sum_{l''=f}^c\sum_{u''=0}^{c-l''}\sum_{v''=0}^{l''}r_{-1}\rho_{-1,l'',u'',v''}
C_{\sigma',l'}(l'',u'',v'')\nonumber\\
&\sum_{\sigma'}\sum_{u'=0}^{c-f}\sum_{v'=0}^{f}P_{{\rm c}1}^{f,\sigma'}(u',v'|-1,l'',u'',v'')
C_{\sigma,l}(l',u',v',-1,l'')\nonumber\\
&[P_{{\rm c}2}^{l,\sigma}(u+\delta(\sigma'+1),v+\delta(\sigma'-1)|(\sigma',f,u',v'),(-1,l'',u'',v''))\nonumber\\
&-P_{{\rm c}2}^{l,\sigma}(u,v|(\sigma',f,u',v'),(-1,l'',u'',v''))],\label{eq2}
\end{eqnarray} where $\rho_{\sigma,luv}\equiv \sum_{i=1}^NP_i(\sigma,l,u,v;t)/N$.
It is plausible that $\rho_{\sigma,luv}$ corresponds to 
$\sum_{i=1}^N\delta(\sigma-\sigma_i)\delta(l-l_i)\delta(u-u_i)\delta(v-v_i)/N$ 
in the thermodynamic limit $N\to\infty$. Using this, the magnetization is expressed as 
$m(t)=\sum_{\sigma}\sum_{l=0}^c\sum_{u=0}^{c-l}\sum_{v=0}^{l}\sigma \rho_{\sigma,luv}(t)$.
In order to obtain a closed dynamical system 
in terms of $\bv{\rho}^{(2)}\equiv(\rho_{\sigma,luv})_{\sigma,luv}$,
 we use the following approximations.
\begin{eqnarray}
P^{l_1,\sigma_1}_{{\rm c}2}(u_1,v_1|(\sigma_2,l_2,u_2,v_2),(\sigma_3,l_3,u_3,v_3))\nonumber\\
= P^{l_1,\sigma_1}_{{\rm c}1}(l_1,u_1,v_1|\sigma_2,l_2\gtreqless f),\label{app21}\\
P^{l,\sigma}_{{\rm c}1}(u,v|\sigma',l',u',v')=
P^{l,\sigma}_{{\rm c}1}(u,v|\sigma',l'\gtreqless f),\label{app22}
\end{eqnarray} which are exact for equilibrium states.
In fact, we can obtain the concrete expression of 
$P_{{\rm c}1}^{l(\gtreqless f),\sigma}(u,v|\sigma',l'{\gtreqless} f)$ as follows.
\begin{eqnarray}
P_{{\rm c}1}^{l(\gtreqless f),\sigma}(u,v|+1,l'\ge f)=(c-l-u)\rho_{\sigma,luv}/
\sum_{l\gtreqless f,uv}(c-l-u)\rho_{\sigma,luv},\label{cal2i}\\
P_{{\rm c}1}^{l(\gtreqless f),\sigma}(u,v|+1,l'<f)=u\rho_{\sigma,luv}/
\sum_{l\gtreqless f,uv}u\rho_{\sigma,luv},\\
P_{{\rm c}1}^{l(\gtreqless f),\sigma}(u,v|-1,l'\ge f)=(l-v)\rho_{\sigma,luv}/
\sum_{l\gtreqless f,uv}(l-v)\rho_{\sigma,luv},\\
P_{{\rm c}1}^{l(\gtreqless f),\sigma}(u,v|+1,l'<f)=v\rho_{\sigma,luv}/
\sum_{l\gtreqless f,uv}v\rho_{\sigma,luv},\label{cal2f}
\end{eqnarray} 
where $\sum_{l\ge f,uv}\equiv \sum_{l=f}^c\sum_{u=0}^{c-l}\sum_{v=0}^l$ 
and $\sum_{l<f,uv}\equiv\sum_{l=0}^{f-1}\sum_{u=0}^{c-l}\sum_{v=0}^l$.
That is, equations (\ref{eq2})$-$(\ref{cal2f}) 
lead to a closed dynamical system 
in terms of $\bv{\rho}^{(2)}$ as follows.
\begin{eqnarray}
\partial_t\bv{\rho}^{(2)}=\bv{G}^{(2)}(\bv{\rho}^{(2)}). \label{dyn2}
\end{eqnarray}

\begin{figure}
\centering
\includegraphics[width=8cm,clip]{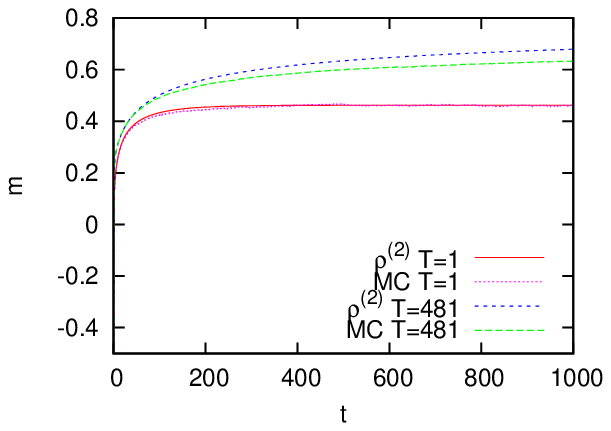}
\caption{Time-dependent magnetization 
$\sum_\sigma\sum_{l=0}^c\sum_{l=0}^{c-l}\sum_{v=0}^l\sigma \rho_{\sigma,luv}$ 
described by (\ref{dyn2}) and 
$\hat{m}$ by the MC simulations with $N=10^6$ for $c=4,k=3$.}
\label{met2}
\end{figure}
The stationary solutions of dynamical system (\ref{dyn2})
also provide exact final states of the system 
above the nonergodic transition point.
Further, the solutions of dynamical system (\ref{dyn2}) 
are closer to the MC simulations than those of previous two descriptions.
However, similar to previous descriptions,  
the solutions are gradually deviated from the MC simulations for lower temperatures, 
as seen in figure \ref{met2}.

\section{On a perturbation series}
\subsection{Systematic improvement of the solutions}
\begin{figure}
\centering
\includegraphics[width=7cm,clip]{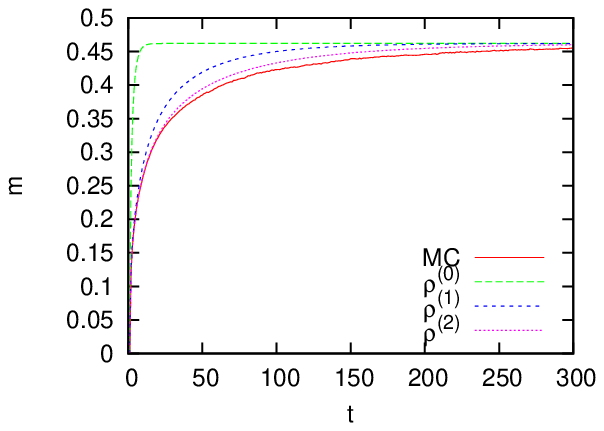}\includegraphics[width=7cm,clip]{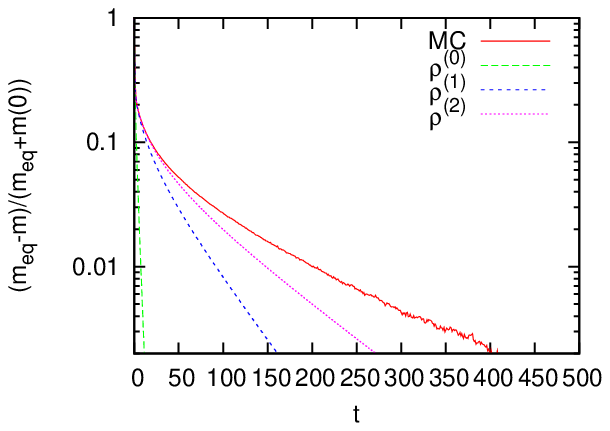}
\caption{The magnetization described by each dynamical systems 
and $\hat{m}$ by the MC simulations at $T=1$. $N=10^7$, $c=4,k=3$.}
\label{compare}
\end{figure}

We consider the manner of changes in the solutions 
of the derived dynamical systems on a perturbation series. 
It is noteworthy that the differences between the MC simulations
and the time-dependent solutions of the derived dynamical systems 
are systematically reduced with the increase 
in the number $n$ of $\bv{\rho}^{(n)}$, 
as observed in the left-hand side of figure \ref{compare}. 

\begin{figure}
\centering
\includegraphics[width=7cm,clip]{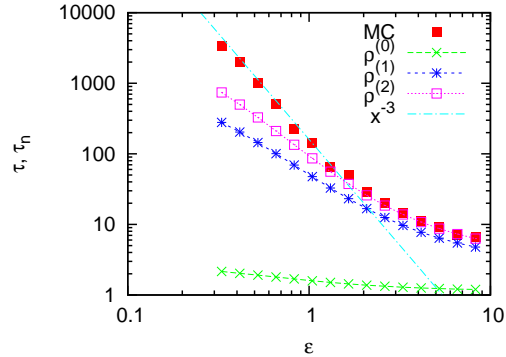}
\caption{$\epsilon$-dependence of each relaxation time. MC simulations 
are performed with $N=10^6$, $c=4,k=3$.}
\label{comparee}
\end{figure}

The MC simulations show that under the initial condition that all spins 
are downward, 
the magnetization behaves as $\hat{m}(t)-m_{\rm eq}(T)\simeq\exp(-t/\tau)$ 
in a long time limit, as shown in the right-hand side of figure \ref{compare}. 
As shown in figure \ref{comparee}, a rough estimation of $\tau$ 
indicates the behavior $\tau\simeq \epsilon^{-\zeta}$, where $\zeta\simeq 3$ 
and $\epsilon\equiv (T/T_{\rm c}-1)$. 
In fact, the value around $\zeta\simeq 3$ has been already confirmed 
for the persistent time in a previous study \cite{Toninelli2}. 

Let us consider the relaxation time $\tau_{n}(T)$ of the dynamical systems 
defined as $|\lambda^{(n)}|^{-1}$, where $\lambda^{(n)}$ is the maximum eigenvalue, 
except for trivial zero, of matrix 
$\mathcal{M}^{(n)}$ obtained by linearizing $\bv{G}^{(n)}$ at a stationary solution 
$\bv{\rho}_{\rm st}^{(n)}\equiv \lim_{t\to \infty} \bv{\rho}^{(n)}(t)$ 
with $\hat{m}(0)=-1$. $\tau_{n}(T)$ does not show 
the power-law behaviour $\epsilon^{-\zeta}$, as seen in figure \ref{comparee} 
although when $\epsilon$ is slightly below $1$, 
$\tau_n(T)$ behaves as $\epsilon^{-\zeta_n}$ with $\zeta_n<\zeta_{n+1}<\zeta$, 
and $\tau_n(T_c)$ is finite.
Thus, higher order analyses will be needed 
for capturing the power-law behavior $\epsilon^{-\zeta}$.

\subsection{Systematic construction of higher order perturbations}
In this section, we consider a perturbation analysis that 
is of a higher order than those discussed thus far. 
Let us reiterate the way to characterize a site discussed in section 4.1.
In the analysis, in order to define the effective state of a target site $i$, 
we use information of sites in set $B_i$.
We regard the way of such a characterization of site $i$ as the first order characterization.
Next, let us reiterate the way to characterize a site discussed in section 4.2. 
In the analysis, in order to define the effective state of a target site $i$, 
we use information of sites in set $B_i$ and set $B_j$ with $j \in B_i$. 
In other words, site $i$ is characterized by the number of sites in set $B_i$, 
where each site in set $B_i$ is characterized 
by the information of the first order characterization.
We regard the way of such a characterization of site $i$ 
as the second order characterization.

Using this second order characterization, 
we can characterize the sites in set $B_i$ 
except for the information of the branch directed to site $i$. 
Next, site $i$ is characterized by the number of sites in set $B_i$, 
where each site in set $B_i$ is characterized by the above characterization.
This procedure defines third order characterization of site $i$.
In the same way, we can define arbitrary 
$n$-th order characterization iteratively.

Further, essentially the same approximation as those used in 
the first order and second order perturbation analysis can be 
applicable to $n$-th order perturbation analysis, 
which provides exact final states of the system as stationary solutions 
above the transition point.
With this procedure, in principle, we can compute the relaxation time $\tau_{n}(T)$ 
described by the dynamical system for all $n$ orders. 
In this perturbation series, the formal `small' parameter can be regarded as 
the distance between which sites are used for 
defining the effective states of a target site.
It should be noted that the concrete value of $n$ 
does not have any significance other than in the sequence,
and $\infty$-order perturbation analysis may provide
the original Master equation in the thermodynamic limit by the definition.

\subsection{A dynamical scaling law in maximum eigenvalues at $T_{\rm c}$}
The systematic improvement of the solutions described 
by the derived dynamical systems 
on a perturbation series 
and the systematic construction of a higher order perturbation series 
motivate us to consider some scaling relation between 
order $n$ in the perturbation series and relaxation time 
(the inverse of the maximum eigenvalue ) $\tau_{n}(T)$.  
The first question raised here is how large $n$ causes 
the divergence of $\tau_{n}(T_{\rm c})$ 
as $\lim_{n\to n_{\rm c}}\tau_{n}(T_{\rm c})\to \infty$.
On the basis of the consideration that 
the size of connected constrained sites 
with the heterogeneous spin configuration resulting 
in the maximum relaxation time
can be infinite for equilibrium spin configurations at $T=T_{\rm c}$, 
we can expect $n_{\rm c}=\infty$. 
Explicitly, using a function $F$ dependent on $(c,k)$, 
we express $\tau_n(T_{\rm c})=F(n,c,k)$ 
where $F(n,c,k)\to\infty$ with $n\to\infty$. 

\begin{figure}
\centering
\includegraphics[width=8cm,clip]{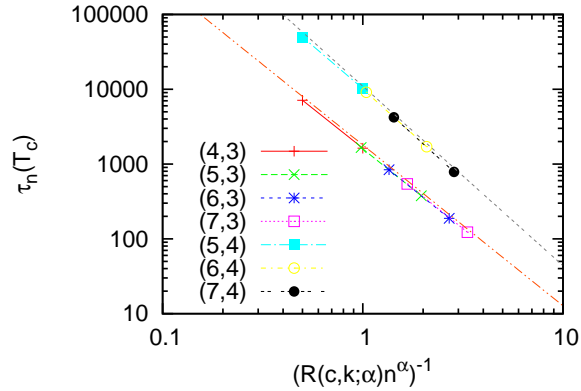}
\caption{$\tau_{n}(T_{\rm c})$ as a function of 
$({\cal R}(c,k;\alpha)n^{\alpha})^{-1}$ 
where we set ${\cal R}(4,3;\alpha)=1$, ${\cal R}(5,4;\alpha)=1$ and $\alpha=1$. 
Fitting parameters are 
${\cal R}(5,3;1)=0.51$, ${\cal R}(6,3;1)=0.37$, ${\cal R}(7,3;1)=0.3$, 
${\cal R}(6,4;1)=0.48$ and ${\cal R}(7,4;1)=0.35$.
$(\cdot,\cdot)$ in the figure is $(c,k)$.}.
\label{scaling}
\end{figure}

As shown in the figure \ref{comparee}, since $\tau_0$ seems 
to be out of the scaling region even if some scaling relation exists,
we focus on $\tau_1(T_{\rm c})$ and $\tau_2(T_{\rm c})$ 
for various parameters $(c,k)$.
Let us remind that this perturbation analysis 
includes the effects of larger connected constrained spins 
if order $n$ is increased.
Therefore, it may be plausible that accessible correlation sizes
by the perturbation analysis with order $n$ are increased 
as order $n$ is increased. 
On the basis of this consideration, first, we assume that 
${\cal R}(c,k;\alpha)n^{\alpha}$ expresses 
accessible correlation sizes by the perturbation analysis with order $n$, 
where ${\cal R}$ and $\alpha$ are certain constants.
It should be noted that the determination of $\alpha$ needs
the information about the nature of correlation sizes in the system, 
which is discussed in section 5.4.
Furthermore, we also assume that $\tau_n$ 
has power-law forms in the accessible correlation sizes in the system. 
Therefore, if we set a value of $\alpha$, 
the above assumptions lead to the exponents of the power-law forms. 
That is, we assume the following form:
\begin{eqnarray}
\tau_n(T_{\rm c})={\cal F}_{\alpha}({\cal R}(c,k ;\alpha)n^{\alpha},c,k), \label{rel}
\end{eqnarray} where ${\cal R}(c,k;\alpha)$ are fitting parameters, 
and ${\cal F}_{\alpha}(x,c,k) = {\cal C} x^{z_*}$.
Surprisingly, as seen in the figure \ref{scaling}, 
we can find that there are values of ${\cal R}$ for arbitrary values of $\alpha$
such that ${\cal F}_{\alpha}(x,c,k)$ is independent of $c$ 
within the numerical analysis.
This result indicates that the assumption for the power-law dependences of 
$\tau_n$ on order $n$ is plausible. 
Concretely, $\alpha z_*(k;\alpha)\simeq z'(k)$ 
$(z'(3)\simeq 2.15),z'(4)\simeq 2.4)$.
In the following, we present some conjectures related to 
the nature of correlation sizes in the system. 

\subsection{Conjectures arising from dynamical scaling law (\ref{rel})}
The law (\ref{rel}) implies that there exists a characteristic size ${\cal N}$ 
obeying the dynamical scaling law $\tau(T)\simeq {\cal N}(T)^{z}$
 where $\tau(T)$ has been already defined through the relaxation 
of the magnetization. 
In fact, we have already known a candidate of ${\cal N}(T)$, which is 
called the {\it minimal rearrangement size} defined as the minimal 
number of flipped spins on the surrounding sites of a target spin 
in order to flip the target spin \cite{Semerjian2,Semerjian3}.
The previous study has captured that the size scale 
shows the power-law behaviour such as $\epsilon^{-\nu(c,k)}$ 
near the nonergodic transition in FA model \cite{Semerjian3}.

Here, we mention to the 
meanings of parameter $\alpha$. 
Rough numerical simulations indicate that 
we need set $\alpha=\alpha_{\rm mrs}\simeq 1.1$ 
in order to identify $z_*$ as $z$.
In this context, $\alpha_{\rm mrs}$ can be regarded as 
the quantity connecting the perturbation order $n$ to 
accessible minimal rearrangement size 
by the perturbation analysis with order $n$. 
In other words, if we obtain a exact value of $\alpha_{\rm mrs}$ such that 
$\alpha_{\rm mrs} z=z'$ in a certain case of $(c,k)=(c',k')$, 
the values of $z(=z'/\alpha_{\rm mrs})$ in any cases $(c,k)\neq (c',k')$ 
can be derived by the perturbation analysis presented above, 
because the analysis provides $z'$ in any cases of $(c,k)$, in principle.

The result for $z_*$ leads to the conjecture that 
$z$ also does not depend on the value of $c$ with the same value of $k$.  
This means that 
the universality classes of nonergodic transitions
in the kinetically constrained spin model 
are classified by the value of constraint parameter $k$, 
and the quantity characterizing the universality 
is not $\zeta$ or $\nu$ but $z$, 
where $\zeta$ and $\nu$ depend on the value of $c$.
Actually, we have performed the MC simulations in order to 
confirm the above conjecture.
Although we have found signs supporting the above conjecture, 
we could not obtain plausible results due to 
the finite size effects and the limitation of the maximum step of time.
It is an important future study to confirm the conjecture by MC simulations.

\section{Concluding remarks}
In this study, we have constructed a systematic perturbation analysis 
for the dynamics of FA model on a Bethe lattice.
This systematic perturbation analysis clarifies the 
existence of a dynamical scaling law, which provides 
a implication  for a universal relation between a size scale 
and a time scale near the nonergodic transition.

Here, we discuss the relevance of our results to the previous studies.
Actually it has been conjectured that the persistent time found by MC simulations 
can be described by a mode-coupling equation \cite{Toninelli2}.
This statement is not inconsistent with $n_{\rm c}=\infty$ 
because mode-coupling equations are $\infty$-dimensional differential equations. 
In addition, a fact supporting the relevance of the model to a mode-coupling equation 
has been made in the literature of the analysis 
of the minimum size rearrangement \cite{Semerjian2}.
In addition to the previous results, 
the results obtained in this paper provide another plausible conjecture for 
the properties of the nonergodic 
transition. That is, the nonergodic transitions have a {\it weak} universality.
This means that critical exponents $\zeta,\nu$ for time and size 
depend on the value of $c$, 
but the dynamical critical exponent $z$ does not depend on 
the value of $c$ with the same value of $k$.
A similar statement has been mentioned in the previous study for 
the dynamical transition in $p$-spin glass model on the Bethe lattice \cite{Semerjian4}. 
The study states that $z$ does not depend on the quantity 
$\gamma$ related to the connectivity of the graph, 
which plays the similar role to that of $c$ in this paper.
However, the study does not mention to the dependence 
of $z$ on the value of $p$, 
which may play the similar role to that of $k$ in this paper. 
The results in this paper indicate that $z$ depends on the value of $p$.
The confirmation of this conjecture for the dynamical transition 
in $p$-spin glass model is an important future study.

Another aspect of this {\it weak} universality appears in the comparison 
to the case of a ferromagnetic Ising model on a Bethe lattice.
In the previous study, the critical exponent can be obtained 
using the dynamical system closed by finite number of effective variables, 
which is derived by the similar approximation method to that of this paper \cite{Semerjian}. 
Therefore, the universality class of nonergodic transition in the KCSM 
is quantitatively and qualitatively different from that of 
ferromagnet-paramagnet transition in some spin models including, at least, 
Ising model. 
Here, the word `qualitatively' means that the differences are located in 
not only the value of critical exponents and the existence of 
the dynamical system closed by finite number of variables 
capturing the critical exponent. 

Finally, we discuss the relationship between nonergodic transitions discussed in this paper 
and the related phase transitions in other systems \cite{Duxbury0}. 
In fact, a decimation dynamics of a random graph in the thermodynamic limit 
exhibits a saddle-node bifurcation at the $k$-core percolation point \cite{Duxbury,Iwata}.
Furthermore, a random-field Ising model 
with zero-temperature Glauber dynamics in the thermodynamic limit 
at a spinodal transition has been reported to correspond to the $k$-core percolation, 
which is also a saddle-node bifurcation on Bethe lattices \cite{Dhar,Ohta1}. 
It should be noted that dynamical behaviours near these transitions corresponding to 
the saddle-node bifurcation are extremely different from the dynamics 
near the nonergodic transition in FA model although those 
are $k$-core percolations from the static viewpoint. 
This difference may be related to the existence of 
a no-passing property in the system \cite{Dhar0}. 
That is, FA model does not have a no-passing property whereas the other systems do.
This consideration leads to the conjecture 
that the nonergodic transition in other kinetically constrained spin models 
such as Kob-Andersen models belong to universality classes 
that are different from that of the saddle-node bifurcation; 
Such models do not have the no-passing property \cite{KA, Toninelli4}. 
Finally, we mention a previous study that suggests a relationship 
between the static properties of jamming transition 
and the $k$-core percolation in a previous study \cite{Schwarz,Schwarz1}.
Since such a system showing the jamming transition do not have 
the no-passing property, 
finding a relationship between the macroscopic dynamical behaviors near 
the jamming transition and KCSM is also an interesting topic for future studies.

\ack
The author wishes to thank S. Sasa for 
providing a critical reading of this paper and related useful feedback, 
G. Semerjian for showing his unpublished note about FA model and
providing many valuable comments, and G. Biroli for providing 
a critical comment; their contributions were immensely beneficial 
in the construction of section 5. 
The author also wishes to thank L. Cugliandolo, R. Jack, K. Miyazaki, 
P. Sollich, and T. Speck for related discussions.
This work was supported by a Grant-in-Aid for JSPS Fellows.

\end{document}